\documentclass
[twocolumn,showpacs,preprintnumbers,prl,tightenlines,
superscriptaddress,footinbib]{revtex4}%
\usepackage{amssymb}
\usepackage{graphicx}
\usepackage{dcolumn}
\usepackage{bm}
\usepackage{amsmath}
\usepackage{amsfonts}%
\providecommand{\U}[1]{\protect\rule{.1in}{.1in}}
\providecommand{\U}[1]{\protect\rule{.1in}{.1in}}
\begin{document}

\title{The entropic origin of disassortativity in complex networks}
\author{Samuel Johnson\thanks{samuel@onsager.ugr.es}, Joaqu\'{\i}n J. Torres,
  J. Marro, and Miguel A. Mu\~{n}oz \\ \small{\textit{Departamento de
      Electromagnetismo y F\'{\i}sica de la Materia, and\\ Institute
      \textit{Carlos I}\ for Theoretical and Computational Physics,\\ Facultad
      de Ciencias, University of Granada, 18071 Granada, Spain.}}}

\begin{abstract}
  Why are most empirical networks, with the prominent exception of
  social ones, generically degree-degree anticorrelated, i.e.
  \textit{disassortative}? With a view to answering this long-standing
  question, we define a general class of degree-degree correlated
  networks and obtain the associated Shannon entropy as a function of
  parameters. It turns out that the maximum entropy does not typically
  correspond to uncorrelated networks, but to either
  \textit{assortative} (correlated) or \textit{disassortative}
  (anticorrelated) ones. More specifically, for highly heterogeneous
  (scale-free) networks, the maximum entropy principle usually leads
  to disassortativity, providing a parsimonious explanation to the
  question above. Furthermore, by comparing the correlations measured
  in some real-world networks with those yielding maximum entropy
for the same degree sequence, 
 we find
a remarkable
agreement in various cases.
  Our approach provides a {\it neutral model} from which, in the absence
  of further knowledge regarding network evolution, one can obtain the expected value of
  correlations. In cases in which empirical observations deviate from
  the neutral predictions -- as happens in social networks -- one can then
  infer that there are specific correlating mechanisms at work.
\end{abstract}

\pacs{89.75.Fb, 89.75.Hc, 05.90.+m}


\maketitle

Complex networks, whether natural or artificial, have non-trivial topologies
which are usually studied by analysing a variety of measures, such as the degree
distribution, clustering, average paths, modularity, etc.
\cite{Nets,Newman_rev,Boccaletti} The mechanisms which lead to a particular
structure and their relation to functional constraints are often not clear
and constitute the subject of much debate \cite{Newman_rev, Boccaletti}. When
nodes are endowed with some additional ``property,'' a feature known as
\textit{mixing} or \textit{assortativity} can arise, whereby edges are not
placed between nodes completely at random, but depending in some way on the property in
question. If similar (dissimilar) nodes tend to wire together, the network is
said to be \textit{assortative} (\textit{disassortative})
\cite{Newman_mixing}.

An interesting situation is when the property taken into account is the degree
of each node -- i.e., the number of neighboring nodes connected to it. It
turns out that a high proportion of empirical networks -- whether biological,
technological, information-related or linguistic -- are disassortatively
arranged (high-degree nodes, or hubs, are preferentially linked to low-degree
neighbors, and viceversa) while social networks are usually assortative. Such
degree-degree correlations have
important consequences for
network characteristics such as connectedness and robustness
\cite{Newman_mixing}.

However, while assortativity in social networks can be explained
taking into account homophily \cite{Newman_mixing} or modularity
\cite{Newman_social}, the widespread prevalence and extent of
disassortative mixing in most other networks remains somewhat
mysterious. Maslov \textit{et al.} found that the restriction of having at most one
edge per pair of nodes induces some disassortative correlations
in heterogeneous networks \cite{Maslov},
and Park and Newman showed how this analogue of the Pauli exclusion principle leads to the edges following Fermi statistics \cite{Park} (see also \cite{Roma}).
However, this
restriction
is not sufficient to fully account for
empirical data. In general,
when one attempts to consider computationally
all the networks with the same distribution as a
given empirical one,
the mean assortativity is not necessarily zero
\cite{Zhao}. But since some ``randomization'' mechanisms induce positive correlations and others negative ones \cite{Farkas}, it is not clear how the phase space can be properly sampled numerically.

In this letter, we show that there is a general reason, consistent
with empirical data, for the ``natural'' mixing of most networks to be
disassortative. Using an information-theory approach we find that the
configuration which can be expected to come about in the absence of specific additional
constraints turns out not to be, in general, uncorrelated. In fact,
for highly heterogeneous degree distributions such as those of the
ubiquitous scale-free networks, we show that the expected value of the
mixing is usually disassortative: there are simply more possible
disassortative configurations than assortative ones. This result
provides a simple topological answer to a long-standing question. Let
us caution that this does {\it not} imply that all scale-free networks
are disassortative, but only that,
in the absence of further information on the mechanisms behind their evolution, this is the neutral expectation.

The topology of a network is entirely described by its adjacency
matrix $\hat{a}$; the element $\hat{a}_{ij}$ represents the number of
edges linking node $i$ to node $j$ (for undirected networks, $\hat{a}$
is symmetric). Among all the possible microscopically distinguishable
configurations a set of $L$ edges can adopt when distributed among $N$
nodes, it is often convenient to consider the set of configurations
which have certain features in common -- typically some macroscopic
magnitude, like the degree distribution. Such a set of configurations
defines an \textit{ensemble}. In a seminal series of papers Bianconi
has determined the partition functions of various ensembles of random
networks and derived their statistical-mechanics entropy
\cite{Bianconi_entropy}. This allows the author to estimate the
probability that a random network with certain constraints has of
belonging to a particular ensemble, and thus assess the relative
importance of different magnitudes and help discern the mechanisms
responsible for a given real-world network. For instance,
she shows
that scale-free networks arise naturally when the total entropy
is restricted to a small finite value. Here we take a similar
approach: we
obtain the Shannon information entropy encoded in the
distribution of edges. As we shall see, both methods yield the same results
\cite{Jaynes}, but for our purposes the Shannon entropy is more tractable.

The Shannon entropy associated with a probability distribution $p_m$ is
$s=-\sum_{m}p_{m}\ln(p_{m})$, where the sum extends over all possible outcomes
$m$. For a given pair of nodes $(i,j)$, $p_{m}$ can be considered to represent
the probability of there being $m$ edges between $i$ and $j$. For simplicity,
we shall focus here on networks such that
$\hat{a}_{ij}$ can only take values $0$ or $1$,
 although the method is applicable to any number of edges allowed. 
In this case, we have only two terms: $p_{1}=\hat{\epsilon}_{ij}$ and
$p_{0}=1-\hat{\epsilon}_{ij}$, where $\hat{\epsilon}_{ij}\equiv
E(\hat{a}_{ij})$ is the expected value of the element $\hat{a}_{ij}$ given
that the network belongs to the ensemble of interest. The entropy associated
with pair $(i,j)$ is then $ s_{ij}=-\left[\hat{\epsilon}_{ij} \ln
 (\hat{\epsilon}_{ij})+(1-\hat{\epsilon}_{ij})\ln(1-\hat{\epsilon}_{ij})\right]
$, while the total entropy of the network is $S=\sum_{ij}^{N}s_{ij}$: 
\begin{equation}
 S=-\sum_{ij}^{N}\left[\hat{\epsilon}_{ij} \ln (\hat{\epsilon}_{ij})
  +(1-\hat{\epsilon}_{ij})\ln(1-\hat{\epsilon}_{ij})\right].
\label{eq_s_exact}
\end{equation}
Since we have not imposed symmetry of the adjacency matrix,
this expression is in general valid for directed networks. For
undirected networks, however, the sum is only over $i\leq j$, with the
consequent reduction in entropy.

For the sake of illustration, we shall estimate the entropy of the Internet at
the autonomous system (AS) level and compare it with the values obtained in
\cite{Bianconi_entropy} assuming the network belongs to two different
ensembles: the fully random graph, or Erd\H{o}s-R\'{e}nyi (ER) ensemble, and the
\textit{configuration} ensemble with a scale-free degree distribution ($p(k)\sim k^{-\gamma}$)
\cite{Newman_rev} and structural cutoff, $k_{i}<\sqrt{\langle k\rangle N}$,
$\forall i$ \cite{Bianconi_entropy} ($\langle k\rangle$ is the mean degree). In this example, we assume the network to
be sparse enough to expand the term $\ln(1-\hat{\epsilon}_{ij})$ in
Eq. (\ref{eq_s_exact}) and keep only linear terms. This reduces
Eq. (\ref{eq_s_exact}) to
$
S_{sparse}\simeq-\sum_{ij}^{N}\hat{\epsilon}_{ij}
[\ln(\hat{\epsilon}_{ij})-1]+O(\hat{\epsilon}_{ij}^{2}).
$
In the ER ensemble, each of $N$ nodes has an equal probability of receiving each of $\frac{1}{2}\langle k\rangle N$ undirected edges.
So, writing $\hat{\epsilon}_{ij}^{ER}=\langle k\rangle/N$, we have
$
S_{ER}=-\frac{1}{2}\langle k\rangle N\left[\ln\left(\langle k\rangle/N\right)-1\right].
$
The configuration ensemble, which imposes a given
degree sequence $(k_{1},... k_{N})$, is defined via the expected value of the
adjacency matrix:
$\hat{\epsilon}_{ij}^{c}=k_{i}k_{j}/(\langle k\rangle
N)$ \cite{Newman_rev,Johnson}. This value 
leads to
$ S_{c}=\langle k\rangle N[\ln(\langle k\rangle
N)+1]- 2N\langle k \ln k\rangle,
$ where $\langle \cdot \rangle \equiv N^{-1}\sum_{i}(\cdot)$ stands for
an average over nodes.
\begin{figure}
[tbh]
\begin{center}
\includegraphics[
height=5cm,
width=7.5cm
]%
{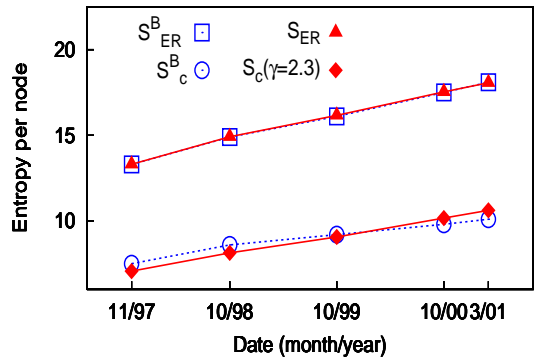}
\caption{(Color online) Evolution of the Internet at the AS
  level. Empty (blue) squares and circles:
entropy per node
  of randomized networks in the fully random and in the configuration ensembles,
  as obtained by Bianconi (hence the ``B'' superscription) \cite{Bianconi_entropy}. Filled
  (red) triangles and diamonds:
Shannon entropy
for an ER network and a scale-free one with $\gamma=2.3$, respectively.
}%
\label{fig_int}%
\end{center}
\end{figure}

Fig. \ref{fig_int} displays the entropy per node obtained in
\cite{Bianconi_entropy} for the first two levels of approximation
(ensembles) to the Internet at the AS level, first
taking into account only the numbers of nodes $N$ and edges $L=\frac{1}{2}\langle
k\rangle N$, and then also the degree sequence. Alongside these, we
plot the 
Shannon entropy
both for an ER random network,
(which coincides exactly with Bianconi's
expression), and for a scale-free network with $\gamma=2.3$
(the slight disparity arising from this exponent's changing a little with time).

We shall now go on to analyse the effect of degree-degree correlations on the
entropy.
In the configuration ensemble, the expected value
of the mean degree of the neighbors of a given node is $
k_{nn,i}=k_{i}^{-1}\sum_{j}\hat{\epsilon}_{ij}^{c}k_{j}=\langle
k^{2}\rangle/\langle k\rangle,
$ which is independent of $k_{i}$. However, as
mentioned above, real networks often display degree-degree correlations, with the result
that $k_{nn,i}=k_{nn}(k_{i})$. If $k_{nn}(k)$ increases (decreases)
with $k$, the network is assortative (disassortative). A measure of
this phenomenon is Pearson's coefficient applied to the
edges \cite{Newman_rev, Newman_mixing, Boccaletti}: $ r= ([
k_{l}k'_{l}]-[ k_{l}]^{2})/([ k_{l}^{2}]-[ k_{l}]^{2}), $ where
$k_{l}$ and $k'_{l}$ are the degrees of each of the two nodes
belonging to edge $l$, and $[\cdot]\equiv(\langle k\rangle
N)^{-1}\sum_{l}(\cdot)$ is an average over edges. Writing
$\sum_{l}(\cdot)=\sum_{ij}\hat{a}_{ij}(\cdot)$, $r$ can be expressed
as
\begin{equation}
  r=\frac{\langle k\rangle \langle k^{2} k_{nn}(k)\rangle - 
    \langle k^{2}\rangle^{2} }{\langle k\rangle \langle k^{3}\rangle 
    - \langle k^{2}\rangle^{2}}.
  \label{r_gen}
\end{equation}
\begin{figure}
[tbh]
\begin{center}
\includegraphics[
height=4.4cm,
width=8.6cm
]%
{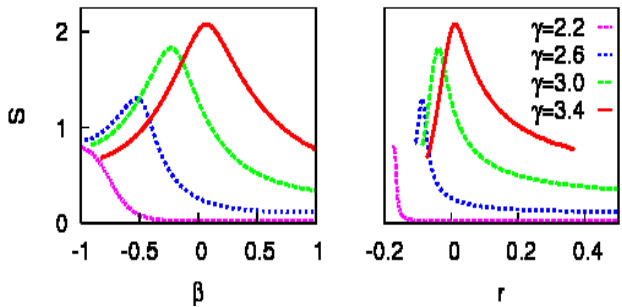}
\caption{(Color online) Shannon entropy of correlated scale-free networks
against parameter $\beta$ (left panel) and
  against Pearson's coefficient $r$ (right panel), for various
  values of $\gamma$ (increasing from bottom to top). $\langle
  k\rangle=10$, $N=10^{4}$.  }%
\label{fig_ent}%
\end{center}
\end{figure}
The ensemble of all networks with a given degree sequence $(k_{1},...k_{N})$
contains a subset for all members of which $k_{nn}(k)$ is constant (the configuration ensemble), but also subsets displaying other functions $k_{nn}(k)$. We can identify each one of these subsets (regions of phase space) with an
expected adjacency matrix $\hat{\epsilon}$ which simultaneously satisfies the following conditions: ${\bf i)}$  $\sum_{j}k_{j}\hat{\epsilon}_{ij}=k_{i}k_{nn}(k_{i})$, $\forall i$, and ${\bf ii)}$ $\sum_{j}\hat{\epsilon}_{ij}=k_{i}$, $\forall i$ (for consistency). An ansatz which fulfills these requirements is any matrix of the form
\begin{equation}
  \hat{\epsilon}_{ij}=\frac{k_{i}k_{j}}{\langle k\rangle N}
  +\int d\nu \frac{f(\nu)}{N}\left[\frac{(k_{i}k_{j})^{\nu}}
    {\langle k^{\nu}\rangle}-k_{i}^{\nu}-k_{j}^{\nu}+\langle k^{\nu}\rangle  \right],
\label{epsi_gen}
\end{equation}
where 
$\nu\in\mathbb{R}$ and the function $f(\nu)$
is in general arbitrary,
although depending on the degree sequence
it shall here be restricted to values
which maintain $\hat{\epsilon}_{ij}\in [0,1]$, $\forall i,j$. This ansatz yields
\begin{eqnarray}
  k_{nn}(k)=\frac{\langle k^{2}\rangle}{\langle k\rangle}
  +\int d\nu f(\nu)\sigma_{\nu+1}\left[\frac{k^{\nu-1}}
    {\langle k^{\nu}\rangle}-\frac{1}{k} \right]
\label{knn_gen}
\end{eqnarray}
(the first term being the result for the
configuration ensemble), where $\sigma_{b+1}\equiv \langle k^{b+1}\rangle -\langle k\rangle
\langle k^{b}\rangle$.
In practice, one could adjust Eq. (\ref{knn_gen}) to fit any
given function $k_{nn}(k)$ and then wire up a network with the desired correlations: it suffices to throw random numbers
according to Eq. (\ref{epsi_gen}) with
$f(\nu)$ as
obtained from
the fit to Eq. (\ref{knn_gen}) \cite{footnote2}. To prove the uniqueness of a matrix $\hat{\epsilon}$ obtained in this way (i.e., that it is the only one compatible with a given $k_{nn}(k)$) assume that there exists another valid matrix $\hat{\epsilon}'\neq\hat{\epsilon}$. Writting $\hat{\epsilon}_{ij}'-\hat{\epsilon}_{ij}\equiv h(k_{i},k_{j})=h_{ij}$, then ${\bf i)}$ implies that $\sum_{j}k_{j}h_{ij}=0$, $\forall i$, while ${\bf ii)}$ means that $\sum_{j}h_{ij}=0$, $\forall i$. It follows that $h_{ij}=0$, $\forall j$.
\\
\linebreak

In many empirical networks, $k_{nn}(k)$ has the form $k_{nn}(k)=A+B
k^{\beta}$, with $A,B>0$ \cite{Boccaletti, Pastor-Satorras} -- the
mixing being assortative (disassortative) if $\beta$ is positive
(negative). Such a case is fitted by Eq. (\ref{knn_gen}) if $f(\nu)=C[\delta(\nu-\beta-1)\sigma_{2}/\sigma_{\beta+2}-\delta(\nu-1)]$, with $C$ a positive constant, since this choice yields
\begin{equation}
  k_{nn}(k)=\frac{\langle k^{2}\rangle}{\langle k\rangle}
  +C\sigma_{2}\left[\frac{k^{\beta}}{\langle
      k^{\beta+1}\rangle}-\frac{1}{\langle k\rangle} \right].
\label{knn_simple}
\end{equation}
After plugging Eq. (\ref{knn_simple}) into Eq. (\ref{r_gen}), one obtains:
\begin{equation}
  r=\frac{C\sigma_{2}}{\langle k^{\beta+1}\rangle}
  \left(\frac{\langle k\rangle \langle k^{\beta+2} \rangle - 
      \langle k^{2}\rangle\langle k^{\beta+1}\rangle }
{\langle k\rangle \langle k^{3}\rangle - \langle k^{2}\rangle^{2}}\right).
\label{r_simple}
\end{equation}
Inserting Eq. (\ref{epsi_gen}) in Eq. (\ref{eq_s_exact}), we can calculate the
entropy of 
correlated
networks
as a function of $\beta$ and $C$ -- or,
by using Eq. (\ref{r_simple}),
as a function of $r$. Particularizing for scale-free networks, then given
$\langle k\rangle$, $N$ and $\gamma$, there is always a certain combination of
parameters $\beta$ and $C$ which maximizes the entropy; we shall call these
$\beta^{*}$ and $C^{*}$.  For $\gamma\lesssim 5/2$ this point corresponds
to $C^{*}=1$. For higher $\gamma$, the entropy can be slightly higher for
larger $C$.
However, for these
values of $\gamma$, the assortativity $r$ of the point of maximum entropy
obtained with $C=1$ differs very little from the one corresponding to
$\beta^{*}$ and $C^{*}$ (data not shown). Therefore, for the sake of
clarity but with very little loss of accuracy, in the following we shall
generically set $C=1$ and vary only $\beta$ in our search for the level
of assortativity, $r^{*}$, that maximizes the entropy given $\langle
k\rangle$, $N$ and $\gamma$.  Note that $C=1$ corresponds to removing the
linear term, proportional to $k_i k_j$, in Eq. (\ref{epsi_gen}), and leaving
the leading non-linearity, $(k_i k_j)^{\beta+1}$, as the dominant one.
\begin{figure}
[tbh]
\begin{center}
\includegraphics[
height=5.5cm,
width=7.5cm
]%
{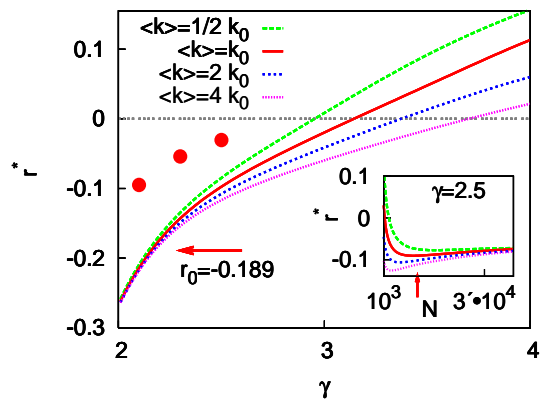}
\caption{(Color online)
Lines from top to bottom: $r$ at which the entropy is maximized, $r^{*}$, against $\gamma$ for random scale-free networks with mean degrees $\langle k\rangle=\frac{1}{2}$, $1$, $2$ and $4$ times $k_{0}=5.981$, and $N=N_{0}=10697$ nodes ($k_{0}$ and $N_{0}$ correspond to the values for the Internet at the AS level in $2001$ \cite{Park}, which had $r=r_{0}=-0.189$). Symbols are the values obtained in \cite{Park} as those expected solely due to the one-edge-per-pair restriction (with $k_{0}$, $N_{0}$ and $\gamma=2.1$, $2.3$ and $2.5$). Inset: $r^{*}$ against $N$ for networks with fixed $\langle k\rangle/N$ (same values as the main panel) and $\gamma=2.5$; the arrow indicates $N=N_{0}$.
}%
\label{fig_max}%
\end{center}
\end{figure}

Fig. \ref{fig_ent} displays the entropy curves for various scale-free
networks, both as functions of $\beta$ and of $r$: {\it depending on
  the value of $\gamma$, the point of maximum entropy can be either
  assortative or disassortative}. This can be seen more clearly in
Fig. \ref{fig_max}, where $r^{*}$ is plotted against $\gamma$ for
scale-free networks with various mean degrees $\langle k\rangle$.
The values obtained by Park and Newman \cite{Park} as those resulting from the one-edge-per-pair restriction are also shown for comparison: notice that whereas this effect alone cannot account for the Internet's correlations for any $\gamma$, entropy considerations would suffice if $\gamma\simeq2.1$.
As shown in the inset, the results are robust in the large system-size limit (although see \cite{Dorogovtsev}).

\begin{figure}
[tbh]
\begin{center}
\includegraphics[
height=5.5cm,
width=7.5cm
]%
{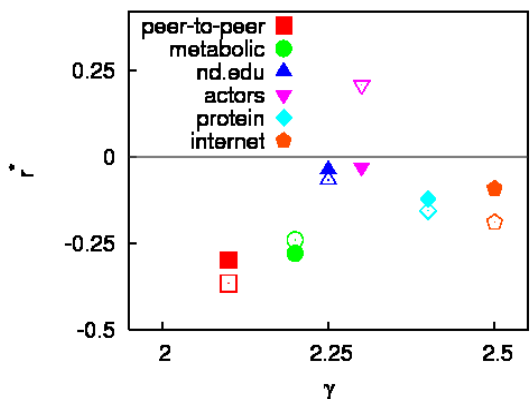}
\caption{(Color online) Level of assortativity that maximizes the
  entropy, $r^{*}$, for various real-world, scale-free networks, as
  predicted theoretically by Eq. (\ref{eq_s_exact}) (solid symbols)
  and as directly measured (empty symbols), against exponent $\gamma$.
}%
\label{fig_all}%
\end{center}
\end{figure}

Since most networks observed in the real world are highly
heterogeneous, with exponents in the range $\gamma\in (2,3)$, it is to
be expected that these should display a certain disassortativity --
the more so the lower $\gamma$ and the higher $\langle k\rangle$. In
Fig. \ref{fig_all} we test this prediction on a sample of empirical,
scale-free networks quoted in Newman's review \cite{Newman_rev} (p. 182). For each case, we found the value of $r$ that maximizes $S$
according to Eq. (\ref{eq_s_exact}), after inserting
Eq. (\ref{epsi_gen}) with the quoted values of $\langle k\rangle$, $N$
and $\gamma$.
In this way, we obtained the expected
assortativity for six networks, representing: a peer-to-peer (P2P)
network, metabolic reactions, the nd.edu domain, actor collaborations,
protein interactions, and the Internet (see \cite{Newman_rev} and
references therein).  For the metabolic, Web domain and protein
networks, {\it the values predicted are in excellent agreement with
  the measured ones}; therefore, no specific anticorrelating mechanisms need to
be invoked to account for their disassortativity. In the other three
cases, however, the predictions are not accurate, so there must be
additional
correlating mechanisms at work. Indeed, it is known that small
routers tend to connect to large ones \cite{Pastor-Satorras}, so one
would expect the Internet to be more disassortative than predicted, as
is the case \cite{footnote3} -- an effect that is less pronounced but still detectable
in the more egalitarian P2P network. Finally, as is typical of social
networks, the actor graph is significantly more assortative than
predicted, probably due to the homophily mechanism whereby highly
connected, big-name actors tend to work together \cite{Newman_mixing}.

In summary, we have shown how
the ensemble of networks with a given degree sequence can be partitioned
into regions of equally correlated networks and found, using an information-theory approach, that
the largest (maximum entropy) region, for the case of scale-free networks, usually displays a certain disassortativity. Therefore, in the absence of knowledge regarding the specific evolutionary forces at work, this should be considered the most likely state.
Given the accuracy with which our approach can predict
the degree of assortativity of certain empirical networks {\it with no
  a priori information thereon}, we suggest this as a neutral model to
decide whether or not particular experimental data require specific
mechanisms to account for observed degree-degree
correlations.

This work was supported by the Junta de Andaluc\'{i}a project
P09-FQM4682, and by Spanish MICINN--FEDER project FIS2009--08451.


\begin{thebibliography}{99}
\bibitem{footnote2} Although, as with the configuration ensemble,
  it is not always possible to wire a network according to a
  given $\hat{\epsilon}$.
\bibitem{footnote3} However, as Fig. \ref{fig_max} shows, if the Internet exponent were the $\gamma=2.2\pm0.1$ reported in \cite{Pastor-Satorras} rather than $\gamma=2.5$, entropy would account more fully for these correlations.
\end{thebibliography}
\end{document}